\documentclass[11pt]{article}
\usepackage{amsmath}
\usepackage{latexsym}
\usepackage{amsfonts}
\usepackage[normalem]{ulem}
\usepackage{soul}
\usepackage{array}
\usepackage{amssymb}
\usepackage{extarrows}
\usepackage{graphicx}
\usepackage[svgnames,table]{xcolor}
\usepackage{tikz}
\usepackage{changepage}
\usepackage{setspace}
\usepackage{ragged2e}
\usetikzlibrary{shapes.symbols,shapes.geometric,shadows,arrows.meta}
\tikzset{>={Latex[width=1.5mm,length=2mm]}}
\usepackage{flowchart}\usepackage[paperheight=11.69in,paperwidth=8.26in,left=0.61in,right=0.61in,top=0.61in,bottom=0.61in,headheight=1in]{geometry}
\usepackage[utf8]{inputenc}
\begin{document}
\pagestyle{empty}

\begin{Center}
{\fontsize{20pt}{36pt}\selectfont \textcolor[HTML]{000000}{Expanding Horizons - Science White Paper}{\fontsize{30pt}{36.0pt}\selectfont \textcolor[HTML]{0070C0}{}\par}\par}

\vspace{3cm}

{\fontsize{20pt}{36pt}\selectfont \textcolor[HTML]{000000}{Exoplanet atmospheres and demographics in the 2040s}{\fontsize{30pt}{36.0pt}\selectfont \textcolor[HTML]{0070C0}{}\par}\par}
\end{Center}\par
\vspace{1cm}

\textbf{Lead author:\\}
Jens Kammerer, European Southern Observatory, Germany\\

\textbf{Co-authors:}\\
Sydney Vach, European Southern Observatory, Germany\\
Sylvestre Lacour, Observatoire de Paris, France\\
Mathias Nowak, Observatoire de Paris, France\\
Thomas Winterhalder, Leiden Observatory, The Netherlands\\
Antoine M\'erand, European Southern Observatory, Germany\\
Akke Corporaal, European Southern Observatory, Chile\\
Guillaume Bourdarot, Max-Planck-Institute for Extraterrestrial Physics, Germany\\
Stefan Kraus, University of Exeter, United Kingdom\\
Sasha Hinkley, University of Exeter, United Kingdom

\newpage

\textbf{Abstract:} Direct observations of exoplanets probe the demographics and atmospheric composition of young self-luminous companions, yielding insight into their formation and early evolution history. In the near future, Gaia will reveal hundreds of nearby young exoplanets amenable to direct follow-up observations. Long-baseline interferometry with current and future facilities is best capable of exploiting this unique synergy which is poised to deliver a statistical sample of benchmark planets with precise dynamical masses and in-depth atmospheric characterization. This will enable tackling the longstanding question of how giant planets form from multiple angles simultaneously, shining light on the complex physical processes underlying planet formation.

\section{Introduction}

To date, more than 6'000 extrasolar planets have been discovered, offering an unprecedented view into the diversity of planetary systems and the outcomes of the planet formation process. However, the vast majority of these planets have been found using indirect detection techniques which predominantly probe mature ($>100$~Myr old) planets and direct insight into the planet formation processes remains elusive. Direct exoplanet detection techniques on the other hand are mostly sensitive to young ($<100$~Myr old) planets  which are still glowing bright in the infrared from their remaining formation heat [1,~2]. Therefore, direct observations of exoplanets are ideally suited to study the physical processes at play during planet formation, and the early dynamical evolution of planetary systems.

Over the past $\sim15$~years, high-contrast imaging instruments on single-dish telescopes have revealed a small but precious population of $\sim30$--40 directly imaged young exoplanets which have been studied in great depth to obtain insights into the planet formation process. Luminosity measurements combined with dynamical masses provide constraints on the post-formation energy budget of young planets, which in turn can reveal the physical processes at play during planet formation [3,~4,~5]. Spectroscopic characterization yields constraints on chemical abundances and and isotopic ratios, which can be traced back to the environment in which the planet formed and migrated [6,~7,~8]. However, the reliable estimation of atmospheric parameters requires the detailed modeling and retrieval of clouds and disequilibrium chemistry in the planet atmosphere, which strongly relies on a broad wavelength coverage [9,~10]. \textbf{In general, the reliability of these various planet formation tracers remains poorly explored through observations. Detailed characterization of a larger sample of planets at a range of ages is required to confirm theoretically predicted population trends in observations.}

More recently, long-baseline interferometry entered the field of exoplanet detection techniques [11]. Observations with VLTI/GRAVITY have since resulted in the characterization of young giant planets with unparalleled astrometric precision and exquisite spectroscopy in the K-band (2.0--2.2~$\text{\textmu m}$) [12,~13,~14,~15]. The GRAVITY data enable the determination of precise dynamical masses if combined with either radial velocity or absolute astrometry data of the host star from Gaia [16], and the atmospheric characterization of substellar objects of L- and T- spectral types through prominent carbon-monoxide (CO) and methane (CH${}_4$) absorption features in the K-band [15].

\section{The VLTI era}

\begin{figure}
    \centering
    \includegraphics[width=\textwidth]{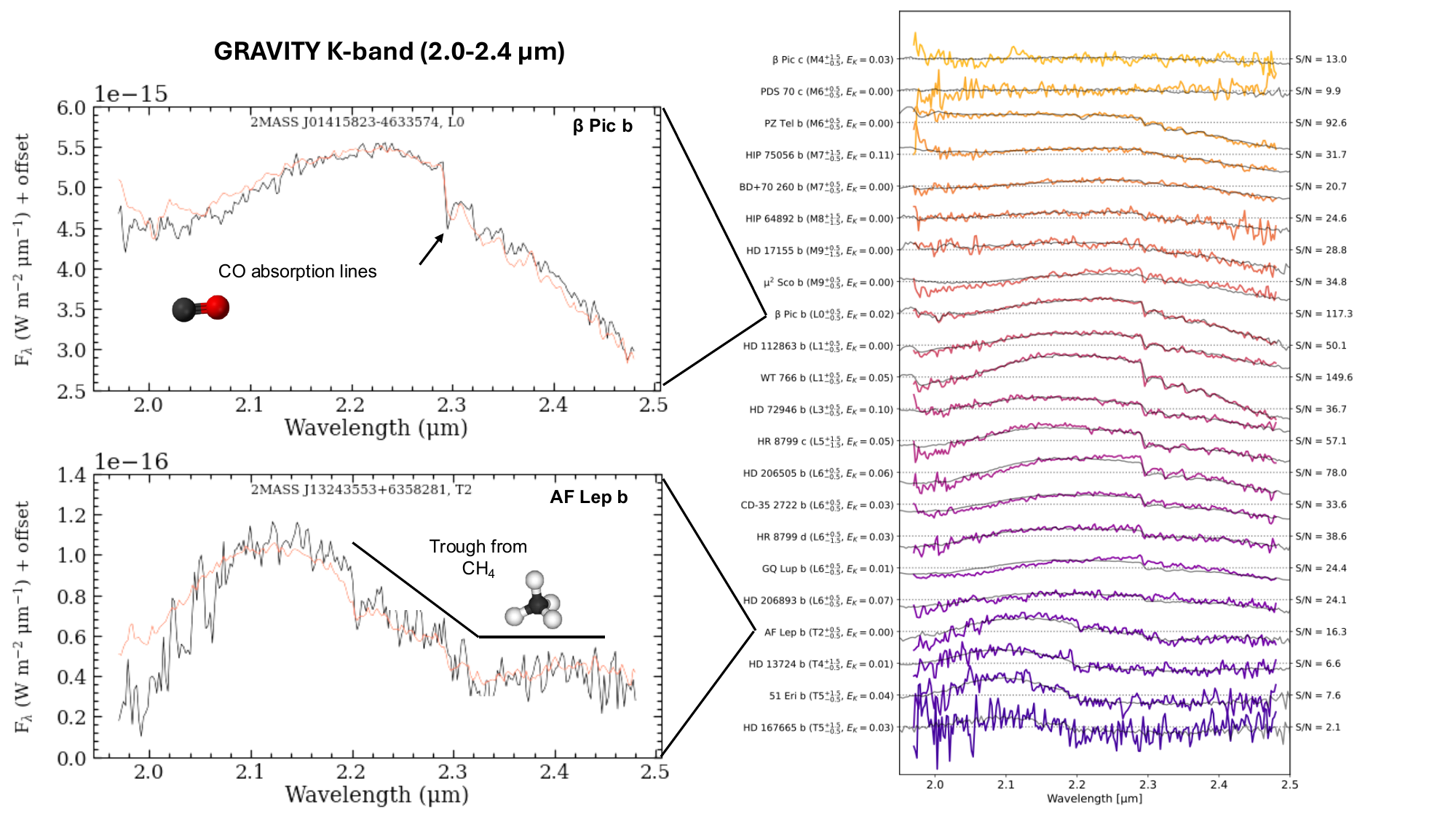}
    \caption{ExoGRAVITY Spectral Library from [15] (right panel) showing the GRAVITY K-band spectra of 22 observed substellar companions sorted by spectral type, from M-type at the top to T-type at the bottom. On the left, zooms on $\beta$~Pic~b (L0 spectral type) and AF~Lep~b (T2 spectral type) are shown, highlighting the chemical features detectable in their spectra.}
    \label{fig:exogravity_spectral_library}
\end{figure}

The unique capability of the VLTI to combine the light of the four 8.2~m Unit Telescopes currently presents the only possibility world-wide to observe exoplanets with long-baseline interferometry. GRAVITY has now characterized a sample of $\sim40$ substellar companions, including $\sim20$ exoplanets [15], and MATISSE has successfully detected the young giant planet $\beta$~Pic~b in the L- and M-bands recently [17]. Interferometric observations with the VLTI provide a unique combination of spectroscopic SNR and angular resolution which is currently beyond the scope of any single-dish telescope. [13] and [14] have demonstrated that GRAVITY is not only capable of characterizing previously known planets, but also of detecting new planets inside the inner working angle of any other currently available high-contrast imaging instrument. At the distance of nearby young moving groups, GRAVITY can observe planets at iceline separations (3--4~au), which corresponds to the separations at which giant planets are predicted to be most common according to formation theories [18] and radial velocity studies [19,~20]. [16] were even able to resolve brown dwarf companions at less than 1~au from the star. Novel fiber mosaicking and speckle nulling techniques now also allow for demographics studies at $\sim100$~mas separation from the star. However, the biggest impact is expected to come from the future release of Gaia DR4 (scheduled for December 2026) which will provide a sizable sample of young giant planet candidates that can be followed up with GRAVITY and MATISSE for in-depth orbital and atmospheric characterization. [15] have shown how exoplanet population studies with interferometry can reveal new insights into the physics at play at the substellar L-T transition. Improved modeling and calibration techniques (Sauter et al. in prep.) will further enhance the photometric accuracy of GRAVITY observations and pave the way for variability studies of substellar companions. [12] have already shown how C/O ratios measured with the help of GRAVITY provide insight into the planet formation history. Stauffenberg et al. in prep. and Spezzano et al. in prep. report the detection of ${}^{13}$CO in young giant planet atmospheres with GRAVITY which can be used to determine isotopic abundance ratios and constrain the environment in which the planet accreted [8]. \textbf{However, the planet sample accessible with current facilities remains limited, prohibiting critically needed demographics studies at iceline separations and limiting detailed atmospheric characterization to a few benchmark objects.}

\section{Expanding horizons}

\begin{figure}
    \centering
    \includegraphics[width=0.6\textwidth]{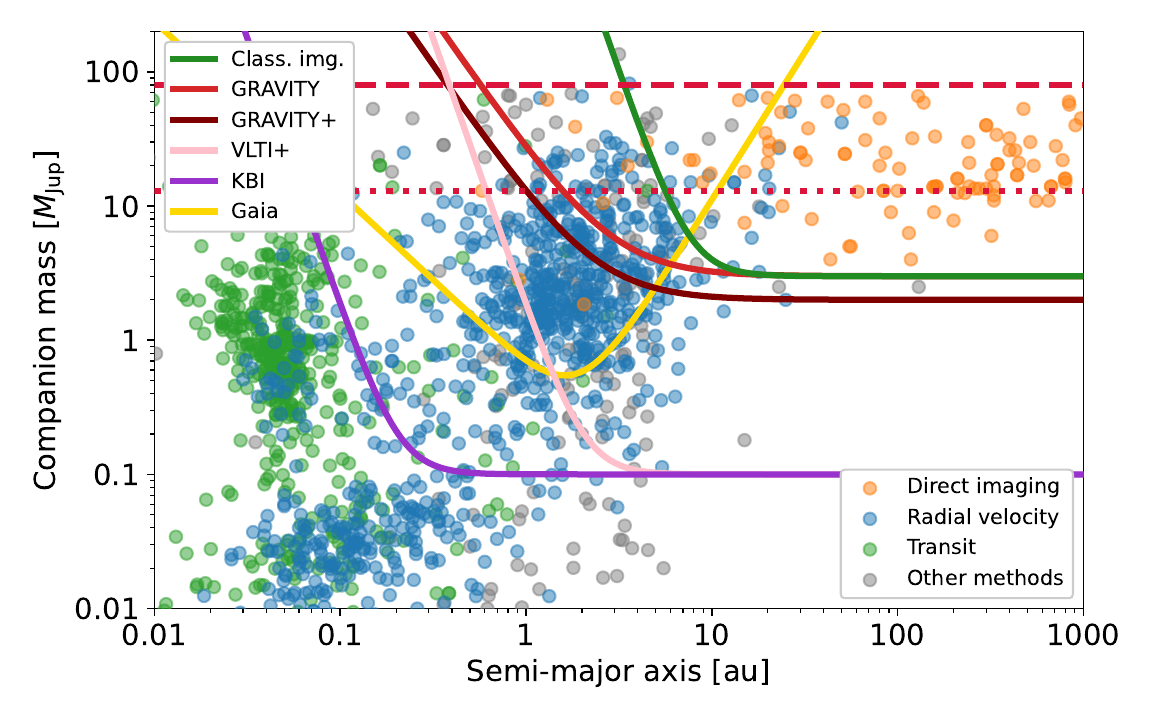}
    \caption{Currently known exoplanet population color-coded by detection technique in the companion mass vs. semi-major axis plane. The detection limits of various facilities are overlaid with solid lines. The boundary between giant planets and brown dwarfs is highlighted by a dotted red line and the boundary between brown dwarfs and stars is highlighted by a dashed red line. VLTI+ is an upgraded VLTI and KPI stands for Kilometric Baseline Interferometry.}
    \label{fig:detection_limits}
\end{figure}

The next generation of optical \& infrared long-baseline interferometers will provide access to a new domain of the exoplanet population. Here, we discuss in particular the expected performance of future dual-field interferometers, building directly on the heritage of GRAVITY and GRAVITY+. As shown in [21], an upgraded VLTI (here referred to as VLTI+) would provide a gain in contrast performance of $\sim10$--100. This would be achieved with improved vibration control, increased throughput, and real-time wavefront control. Moreover, by adding a fifth Unit Telescope with a baseline of $\sim200$~m to the VLTI, the angular resolution can also be boosted, enabling the detection of mature planets in reflected starlight. \textbf{This would allow the measurement of giant planet geometric albedo as a function of wavelength which provides key insight into the atmospheric composition and cloud properties of these objects.} Such a facility would complement the Roman Space Telescope, which will be capable of observing a handful of giant planets in reflected light too, but at much larger angular separation from the star. Together, these pioneering observations will lay the foundations for the direct observation of rocky (potentially habitable) planets in reflected starlight with the Habitable Worlds Observatory.

For young planets in nearby moving groups, a VLTI+ type facility would dramatically increase the overlap with the astrometric planet discovery space of the Gaia satellite. The scientific power of this overlap has already been demonstrated by [16] who measured precise dynamical masses for five substellar companions by combining astrometry from Gaia with a single epoch of GRAVITY data. \textbf{The determination of model-independent dynamical masses, enabled for a population-size sample of planets by a VLTI+ type facility, is key for multiple aspects of planet characterization.} Firstly, they are crucial for constraining atmospheric retrievals which otherwise suffer from strong correlations between surface gravity, metallicity, clouds, and disequilibrium chemistry. Secondly, combined with planet luminosity measurements they can constrain the thermal energy budget of the planet, providing insights into its formation mechanism. Thirdly, dynamical masses and the precise astrometry from interferometry allow for a detailed study of planet orbits and their evolution. For young planets embedded in disks, these can be compared to hydrodynamical simulations of the disk to constrain formation mechanism and disk viscosity. For more mature planets who are no longer embedded, planet spin from e.g., planetary radial velocity measurements can be compared to orbital axis to constrain planet obliquity, which also provides clues about the formation history.

An even larger angular resolution could be achieved by adding telescopes with kilometric baselines to the VLTI. Such a facility, here referred to as kilometric baseline interferometry or KBI, would enable the detection of the bulk of the giant planet population at $\sim1$--4~au at larger distances, out to the most nearby star-forming regions. \textbf{This would enable transformative studies of the youngest giant planet atmospheres, still carrying the most pristine chemical imprint of the planet formation process.} Comparison to the atmospheres of the more mature young moving group objects would inform about early-on atmospheric evolution and the reliability of chemical formation tracers at intermediate ages. Such observations are beyond the reach even of the ELT. The greatly enhanced angular resolution of a KBI-type facility allows for superior starlight rejection at close angular separations, yielding higher signal-to-noise exoplanet spectra, at a higher spectral resolution.

{\tiny \singlespacing
\section{References}
[1] B. P. Bowler (2016), PASP, 128, 2001,
[2] T. Currie et al. (2023), ASPC, 534, 799,
[3] D. S. Spiegel \& A. Burrows (2012), ApJ, 745, 174,
[4] C. Mordasini et al. (2012) , A\&A, 547, 111,
[5] G.-D. Marleau \& A. Cumming (2014), MNRAS, 437, 1378,
[6] K. I. \"Oberg et al. (2011), ApJ, 743, 16,
[7] C. Eistrup et al. (2018), A\&A, 613, 14,
[8] Y. Zhang et al. (2021), Nature, 595, 370,
[9] B. E. Miles et al. (2023), ApJ, 946, 6,
[10] K. K. W. Hoch et al. (2024), vol. 168, p. 187, Oct. 2024.
[11] GRAVITY Collaboration et al. (2019), A\&A, 623, 11,
[12] GRAVITY Collaboration et al. (2020), A\&A, 633, 110,
[13] M. Nowak et al. (2020), 642, 2,
[14] S. Hinkley et al. (2023), A\&A, 671, 5,
[15] J. Kammerer et al. (2025), arXiv e-prints, p. arXiv:2510.08691,
[16] T. O. Winterhalder et al. (2024), A\&A, 688, 44,
[17] M. Houll´e et al. (2025), arXiv e-prints, p. arXiv:2508.18366,
[18] G. D’Angelo et al. (2010), Exoplanets, 319,
[19] R. B. Fernandes et al. (2019), ApJ, 874, 81,
[20] B. J. Fulton et al. (2021), ApJS, 255, 14,
[21] S. Lacour et al. (2025), A\&A, 694, 277.
}

\newpage


\end{document}